\documentclass[11pt]{article}
\usepackage{graphicx}
\usepackage{amsmath}
\usepackage{amssymb}
\usepackage{amsxtra}
\usepackage{changebar}
\usepackage{placeins}

\newcommand{\p}{\partial}

\newcommand{\beq}{\begin{equation}}
\newcommand{\eeq}{\end{equation}}
\newcommand{\nin}{\noindent}

\newcommand{\bz}{\mbox{\bf{Z}}}

\newcommand{\Rie}{R_{\mu\nu\rho}^{\;\;\;\;\;\;\;
\sigma}}

\begin{document}

\title{Explaining Phenomenologically Observed Spacetime Flatness Requires New Fundamental Scale Physics}

\author{D. L. Bennett\\ \small{Brookes Institute for Advanced Studies, B{\o}gevej 6, 2900 Hellerup,
Denmark} \\ \small{dlbennett99@gmail.com}\\ H. B. Nielsen \\
\small{The Niels Bohr Institute, Blegdamsvej 17, 2100
Copenhagen {\O }}
\\ \small{hbech@nbi.dk}}

\maketitle

\begin{abstract}
The phenomenologically observed flatness - or near flatness -
of spacetime cannot be understood as emerging from continuum
Planck (or sub-Planck) scales using known physics. Using
dimensional arguments it is demonstrated that any immaginable
action will lead to Christoffel symbols that are chaotic. We
put forward new physics in the form of fundamental fields that
spontaneously break translational invariance. Using these new
fields as coordinates we define the metric in such a way that
the Riemann tensor vanishes identically as a Bianchi identity.
Hence the new fundamental fields define a flat space. General
relativity with curvature is  recovered as an effective theory
at larger scales at which crystal defects in the form of
disclinations come into play as the sources of curvature.

\end{abstract}

\newpage


\vspace{1cm}

\section{Introduction}
\nin We address the fundamental mystery of why the spacetime
that we experience in our everyday lives is so nearly flat.
More provocatively one could ask why the macroscopic spacetime
in which we are immersed doesn't consist of spacetime
foam\cite{misner,wheeler}.

This question is approached by putting up a NO-GO for having
the spacetime flatness that we observe phenomenologically. This
NO-GO builds upon an argumentation that starts with the
assumption that spacetime is a continuum down to arbitrarily
small scales $a$ with $a<<l_{Pl}$ where $l_{Pl}$ is the Planck
length.

Earlier one of us (H.B.N.) has attempted to derive
reparametrization invariance as a consequence of quantum
fluctuations \cite{lehto}. If reparametrization invariance were
for such a reason exact, it would be difficult to see how
accepting arbitrarily small length scales could be avoided. So
that would nessesitate our assumtion of a total continuum.''

This assumption of a continuum at all scales $a<<l_{Pl}$
forbids having any form of regulator - e.g., a lattice. With no
regulator in place we must expect enormous quantum fluctuations
unless we can come to think of some physics that can tame them.

We shall argue quite generally that no form of known physics
can accomplish this. For example, the Einstein-Hilbert action

\beq \frac{1}{2\kappa}\int d^4 x \sqrt{-g} R \eeq

\nin hasn't a chance since at scales $a$ for which $
\frac{1}{a^2 }>> \frac{1}{\kappa}$ this action is negligible.
We shall argue that there does not exist a functional form for
an action that can prevent spacetime foam for arbitrarily small
scales $a<<l_{Pl}$.

As a solution to this problem we propose new fundamental fields
at scales $a<<l_{Pl}$ that spontaneously break translational
invariance. This approach was inspired by the
work\cite{Guendelman} of Eduardo Guendelman.

\section{Phenomenological Flatness Impossible if Spacetime Foam Shows Up at Any Scale Including Scales $a<<l_{Pl}$}

Over long distances the spacetime that we experience is -
barring the presence of nearby gravitational singularities -
very nearly flat. This means that the parallel transport of a
vector from a spacetime point $A$ to a distant spacetime point
$B$ along say many different pathes should result in a
well-defined (small) average rotation angle for the parallel
transported vector.

If the connection used for parallel transport takes values in a
compact group, a path along which there is strong curvature can
have an orbit on the group manifold that is wrapped around the
group manifold several or many times depending on the amount of
curvature.

Take an $S^1$ as a prototype compact group manifold. The
rotation under parallel transport can be written

\beq \Theta = \theta + 2\pi k \eeq

\nin $-\pi < \theta < \pi$ and $k \in \bz $.

For nearly flat space the rotation angle $\Theta $ under
parallel transport along a path will vary very slowly along the
path. The average values of $\Theta$ along different pathes are
expected to be closely clustered around $ \Theta = 0$ and with
certainty to lie in the interval $[-\pi, \pi]$.

However, if there were an underlying spacetime foam, then  two
pathes $\Gamma_1$ and $\Gamma_2$ connecting the same two widely
separated points would in general have vastly different values
of $\Theta$ say $\Theta_1$ and $\Theta_2$ reflecting the fact
that the enormous curvatures encountered in traversing the
spacetime foam along the two pathes are completely
uncorrelated. If

\beq \Theta_1 = \theta_1 + 2\pi k_1 \eeq

\nin and

\beq \Theta_2 = \theta_2 + 2\pi k_2 \eeq

\nin we expect $k_1$ and $k_2$ to be  large and uncorrelated
which also means that $\theta_1$ and $\theta_2$ are completely
uncorrelated as to their position in the interval $[-\pi,
\pi]$.

For example, the pathes $\Gamma_1$ and $\Gamma_2$ could have
$\Theta_1$ and $\Theta_2$ values such that $k_2 >> k_1 >> 1$
while $\Theta_1$ mod $2\pi = \theta_1$ and $\Theta_2$ mod $2\pi
= \theta_2$ could be such that $\theta_1 > \theta_2$. So it
does not necessarily follow from  $\Theta_2 >\Theta_1$  that
$\theta_2 > \theta_1$.

The fact that $\Theta$ and $\theta$ can differ by a term that
is the product of a uncontrollably large number $|k_2-k_1|$
multiplied by $2\pi$ means that the idea of an average rotation
angle when parallel transporting a vector along different
pathes between two spacetime points is meaningless. The
underlying reason is that in traversing spacetime foam the
connection is an uncontrollably rapidly varying function of any
path going through spacetime foam. In particular this argument
would also apply to pathes connecting spacetime points
separated by distances for which spacetime is known
phenomenologically to be flat or at least nearly flat.

Without a well defined connection the concept of spacetime
flatness is meaningless. The conclusion is that if spacetime
foam comes into existence at {\it any} scale under the scale at
which we phenomenologically observe flatness, the possibility
for having flat spacetime is forever lost.

It is in particular at scales $a$ with $a << l_{Pl}$ that there
is the danger of spacetime foam coming into existence. At these
scales the Einstein-Hilbert action would be completely
ineffective in preventing spacetime foam. This is the reason
for our proposal of new physics at sub-Planckian scales in the
form of fundamental fields $\phi^a$ (that we also  call
Guendelmann fields) that spontaneously break translational
invariace in the vacuum in such a way that the metric can be
defined by

\beq g_{\mu\nu}= \frac{\p \phi^a \p \phi^b}{\p x^{\mu}\p
x^{\nu}}\eta_{ab}.\label{pivotal}\eeq

\nin With the fundamental fields $\phi^{a}$ defined by this
form for the metric $g_{\mu\nu}$ it can be shown that the
Riemann curvature vanishes identically. The converse  can also
be shown: the condition $R_{\mu\nu\rho}^{\;\;\;\;\;\;\;
\sigma}=0$ implies that $g_{\mu\nu}$ must have the form of
Eqn.(\ref{pivotal}). It should be stressed that $g_{\mu\nu}$
with the form of Eqn.(\ref{pivotal}) leads to
$R_{\mu\nu\rho}^{\;\;\;\;\;\;\; \sigma}=0$ as an identity quite
independently of any choice of Lagrangian (or lack thereof) and
the equations of motion that follow from such a choice.

\section{There Exists No Action Depending Only on Translationally Invariant Coordinates that can Keep Spacetime
Flat at All Scales}

We consider the variation of the rotation angle of a vector
field (or in general a tensor field) parallel transported
around a loop of radius $a$ as $a$ goes to values much Less
than the Planck scale compared say to the angle $2\pi$. For
this purpose we consider the connection
$\Gamma_{\mu\nu}^{\;\;\;\;\;\rho}$ integrated around the edge
of a disc of radius $a$:

\beq \oint_{\mbox{\tiny{disc edge }}2\pi
a}\Gamma_{\mu\nu}^{\;\;\;\;\;\rho}dx^{\nu}\stackrel{Stokes}{\approx}\int_{\mbox{\tiny{disc
area }}\pi
a^2}R_{\mu\nu\lambda}^{\;\;\;\;\;\rho}dx^{\nu}dx^{\lambda}\eeq

\section{Flatness Requires New Fundamental Fields that Break Translational Invariance Spontaneously at Sub-Planck Scales}

We introduce new fundamental fields $\phi^a(x^{\mu})$ at scales
$a$ with $a<<l_{Pl}$ that spontaneously break translational
invariance in such a way that the metric is defined by

\beq g_{\mu\nu}= \frac{\p \phi^a \p \phi^b}{\p x^{\mu}\p
x^{\nu}}\eta_{ab}.\label{pivotal1}\eeq

The new fundamental fields can also be thought of as
fundamental absolute coordinates insofar as they break
translational invariance. By indexing the new fields
(coordinates)  $\phi^a$ with indices $a,b,c,...$ we are
anticipating a later development in which these indices will be
seen to be flat indices.

At this point we shall show explicitly the important property
that the Riemann tensor $\Rie$ vanishes identically when the
new fundamental coordinates $\phi^a,\phi^b,\phi^c,...$ are
chosen as in Eqn. (\ref{pivotal}). To this end we need
Christoffel symbols

$$\Gamma^{\rho}_{\;\;\mu\nu}=\frac{1}{2}g^{\rho\tau}\left( \frac{\p g_{\nu\tau}}{\p x^{\mu}}+
\frac{\p g_{\mu\tau}}{\p x^{\nu}}-
\frac{\p g_{\mu\nu}}{\p x^{\tau}}  \right)$$

\nin in the form

$$\Gamma_{\gamma\mu\nu}=g_{\gamma\rho}\Gamma^{\rho}_{\;\;\mu\nu}=\frac{1}{2}\left(\frac{\p g_{\nu\gamma}}{\p x^{\mu}}+
\frac{\p g_{\mu\gamma}}{\p x^{\nu}}-\frac{\p g_{\mu\nu}}{\p x^{\gamma}}\right)$$

\nin into which we substitute Eqn. (\ref{pivotal})

$$ =\frac{1}{2}\eta_{ab}\left(\frac{\p}{\p x^{\mu}}(\frac{\p \phi^a}{\p x^{\nu}}\frac{\p \phi^b}{\p x^{\gamma}})+
\frac{\p}{\p x^{\nu}}(\frac{\p \phi^a}{\p x^{\mu}}\frac{\p \phi^b}{\p x^{\gamma}})-\frac{\p}{\p x^{\gamma}}(\frac{\p \phi^a}{\p x^{\mu}}\frac{\p
\phi^b}{\p x^{\nu}})\right)$$

\nin which reduces to

$$\eta_{ab}\frac{\p^2 \phi^a}{\p x^{\mu}\p x^{\nu}}\frac{\p \phi^b}{\p
x^{\sigma}}.
$$

\nin Going from $\Gamma_{\gamma\mu\nu}$ back to
$\Gamma^{\rho}_{\;\;\mu\nu}=g^{\gamma\rho}\Gamma_{\gamma\mu\nu}$
yields

\beq\Gamma^{\rho}_{\;\;\mu\nu}=\eta_{ab} g^{\rho\sigma}\frac{\p
\phi^b}{\p x^{\sigma}}\frac{\p^2 \phi^a}{\p x^{\mu}\p
x^{\nu}}.\label{back}\eeq

\nin We want to show that the the Riemann tensor

$$
R^{\;\;\;\;\;\;\;\sigma}_{\mu\nu\lambda}\doteq
\p_{\mu}\Gamma^{\;\;\;\sigma}_{\nu\lambda}-
\p_{\nu}\Gamma^{\;\;\;\sigma}_{\mu\lambda}-
\Gamma^{\;\;\;\delta}_{\mu\lambda}\Gamma^{\;\;\;\sigma}_{\nu\delta}+
\Gamma^{\;\;\;\delta}_{\nu\lambda}\Gamma^{\;\;\;\sigma}_{\mu\delta}
$$

\nin vanishes identically with the choice Eqn. (\ref{pivotal})
for $g_{\mu\nu}$.

We make a small digression in order to establish an
intermediate result. Consider the matrix element

$$ [g]^{\mu\nu}=\left[(g_{\bullet
\bullet})^{-1}\right]^{\mu\nu}=\left[\left(\frac{\p \phi^a}{\p\
x^{\bullet}}\eta_{ab}\frac{\p \phi^b}{\p
x^{\bullet}}\right)^{-1}\right]^{\mu\nu}=\left[\left(\frac{\p
\phi^{\circ}}{\p
x^{\bullet}}\right)^{-1}\eta_{\circ\circ}^{-1}\left(\frac{\p
\phi^{\circ}}{\p x^{\bullet}}\right)^{-1} \right]^{\mu\nu}
$$\beq =\left[\left(\frac{\p \phi^{\circ}}{\p
x^{\bullet}}\right)^{-1}\right]^{\mu}_{\;a}
\left[\left(\eta_{\bullet\bullet}\right)^{-1}\right]^{ab}
\left[\left(\frac{\p \phi^{\circ}}{\p
x^{\bullet}}\right)^{-1}\right]_b^{\;\nu} \label{matrix} \eeq

\nin where square brackets denote matrix elements with row
indices to the left and column indices to the right. The
symbols $\bullet$ and $\circ$ stand for respectively general
coordinate and flat coordinate indices and are used to indicate
the number and position of otherwise unspecified indices.

Converting from matrix element notation to operator notation
according to

\beq \left[\left(\frac{\p \phi^{\circ}}{\p
x^{\bullet}}\right)^{-1}\right]^{\mu}_{\;a} = \frac{\p
x^{\mu}}{\p \phi^a} \eeq


\nin we have

\beq g^{\mu\nu}=\frac{\p x^{\mu}}{\p \phi^a}
\eta^{-1\;ab}\frac{\p x^{\nu} }{\p \phi^b}.\eeq

\nin multiplying $g^{\mu\nu}$ into $\frac{\p \phi^c}{\p
x^{\nu}}\eta_{bc}$ gives

\beq g^{\mu\nu}\frac{\p \phi^c}{\p x^{\nu}}\eta_{bc} = \frac{\p
x^{\mu}}{\p \phi^a} \eta^{-1\;ad}\frac{\p x^{\nu} }{\p
\phi^d}\frac{\p \phi^c}{\p x^{\nu}}\eta_{bc}=\frac{\p
x^{\mu}}{\p \phi^a}\frac{\p x^{\nu}}{\p\phi^a}\frac{\p
\phi^b}{\p x^{\nu}}= \frac{\p x^{\mu}}{\p \phi^a}\delta^a_{\;
b} =\frac{\p x^{\mu}}{\p \phi^b}\eeq


\nin multiply now both sides of (\ref{back}) by $\frac{\p
\phi^c}{\p x^{\rho}}$

$$\frac{\p \phi^c}{\p
x^{\rho}}\Gamma^{\rho}_{\;\;\mu\nu}=\frac{\p \phi^c}{\p
x^{\rho}}\underbrace{\eta_{ab} g^{\rho\sigma}\frac{\p \phi^b}{\p
x^{\sigma}}}_{ \p x^{\rho}/\p \phi^a \mbox{ {\small from} } (\ref{back})}\frac{\p^2 \phi^a}{\p x^{\mu}\p x^{\nu}} =
\frac{\p^2 \phi^c}{\p x^{\mu}\p x^{\nu}}\label{intermediate}$$

\nin which is the intermediate result needed below.

To show that the Riemann $\Rie$ tensor vanishes identically
when $g^{\mu\nu}$ is chosen to have the form of Eqn.
(\ref{pivotal}) we shall show that

$$\frac{\p \phi^b}{\p x^{\sigma}}R^{\;\;\;\;\sigma}_{\mu\nu\lambda}\equiv 0 $$

\nin for arbitrary $\p \phi^b/\p x^{\sigma}$. Explicitely

$$\frac{\p \phi^b}{\p x^{\sigma}}R^{\;\;\;\;\;\;\;\sigma}_{\mu\nu\lambda}=
\frac{\p \phi^b}{\p
x^{\sigma}}\p_{\mu}\Gamma^{\;\;\;\sigma}_{\nu\lambda}- \frac{\p
\phi^b}{\p
x^{\sigma}}\p_{\nu}\Gamma^{\;\;\;\sigma}_{\mu\lambda}- \frac{\p
\phi^b}{\p
x^{\sigma}}\Gamma^{\;\;\;\delta}_{\mu\lambda}\Gamma^{\;\;\;\sigma}_{\nu\delta}+
\frac{\p \phi^b}{\p
x^{\sigma}}\Gamma^{\;\;\;\delta}_{\nu\lambda}\Gamma^{\;\;\;\sigma}_{\mu\delta}.
\label{derRie} $$

\nin The first two terms on the right hand side, i.e.,

$$\frac{\p \phi^b}{\p
x^{\sigma}}\p_{\mu}\Gamma^{\;\;\;\sigma}_{\nu\lambda}- \frac{\p
\phi^b}{\p
x^{\sigma}}\p_{\nu}\Gamma^{\;\;\;\sigma}_{\mu\lambda}$$

\nin can be written as

$$\frac{\p}{\p x^{\mu}}\left( \frac{\p \phi^b}{\p
x^{\sigma}}\Gamma^{\;\;\;\sigma}_{\nu\lambda}\right)-\frac{\p^2
\phi^b}{\p x^{\mu}\p x^{\sigma}}\Gamma^{\;\;\;\sigma}_{\nu\lambda}
-\left[ \frac{\p}{\p x^{\nu}}\left( \frac{\p \phi^b}{\p
x^{\sigma}}\Gamma^{\;\;\;\sigma}_{\mu\lambda}\right)-\frac{\p^2
\phi^b}{\p x^{\nu}\p x^{\sigma}}\Gamma^{\;\;\;\sigma}_{\mu\lambda}
\right].$$

\nin Using (\ref{intermediate}) to rewrite the 1st and 3rd
terms of this expression gives

$$\frac{\p}{\p x^{\mu}}\frac{\p^2\phi^b}{\p x^{\nu}\p x^{\lambda}}-\frac{\p^2
\phi^b}{\p x^{\mu}\p x^{\sigma}}\Gamma^{\;\;\;\sigma}_{\nu\lambda}-
\frac{\p}{\p x^{\nu}}\frac{\p^2\phi^b}{\p x^{\mu}\p x^{\lambda}}+
\frac{\p^2
\phi^b}{\p x^{\nu}\p x^{\sigma}}\Gamma^{\;\;\;\sigma}_{\mu\lambda}.
 $$
\nin The 1st and 3rd terms cancel since they are totally
symmetric under permutations of the indices $\mu\nu\lambda$.
Consequently what remains of the first two terms of $(\p
\phi^b/\p x^{\sigma})R^{\;\;\;\;\;\sigma}_{\mu\nu\lambda}$ is

$$\frac{\p^2
\phi^b}{\p x^{\nu}\p x^{\sigma}}\Gamma^{\;\;\;\sigma}_{\mu\lambda}- \frac{\p^2
\phi^b}{\p x^{\mu}\p x^{\sigma}}\Gamma^{\;\;\;\sigma}_{\nu\lambda}.  $$

\nin Using the intermediate result (\ref{intermediate}) in
reverse these first two terms of $(\p \phi^b/\p
x^{\sigma})R^{\;\;\;\;\;\sigma}_{\mu\nu\lambda}$ become

$$\frac{\p\phi^b}{\p x^{\rho}}\Gamma^{\;\;\;\sigma}_{\mu\lambda}\Gamma^{\;\;\;\sigma}_{\nu\sigma}-
\frac{\p\phi^b}{\p x^{\rho}}\Gamma^{\;\;\;\sigma}_{\nu\lambda}\Gamma^{\;\;\;\rho}_{\mu\sigma}$$

\nin which are seen to cancel the last two terms of $(\p
\phi^b/\p x^{\sigma})R^{\;\;\;\;\;\sigma}_{\mu\nu\lambda}$ (see
(\ref{derRie}) above). As $(\p \phi^b/\p
x^{\sigma})R^{\;\;\;\;\;\;\sigma}_{\mu\nu\lambda}$ vanishes
identically for arbitrary $\frac{\p \phi^b}{\p x^{\sigma}}$ we
can conclude that the Riemann curvature
$R^{\;\;\;\;\;\;\sigma}_{\mu\nu\lambda}$ vanishes identically
when the metric has the form of Eqn. (\ref{pivotal}). This
result is not surprising in light of our not having introduced
any gravitational sources at the scales $a$ ($a<<l_{Pl}$) at
which we postulate that the new fundamental fields are
instrumental in preventing spacetime foam.

\section{Using the Postulated Fundamental Fields $\phi^a$ to
Build Flat Spacetime: a Model}

Here we put up a model with a term in the Lagrange density
$\mathcal{L}$ that depends on the gradient of the new
fundamental fields $\phi^a,\phi^b,...$.

\beq
\mathcal{L}=\mathcal{L}(\cdots,\phi^{a},\phi^b,\cdots\partial_{\mu}\phi^a,
\partial_{\nu}\phi^b,\cdots)=
\cdots +
\left(t_{ab}(g_{\mu\nu}\partial_{\mu}\phi^a\partial_{\nu}\phi^b-p
\eta^{ab})\right)^2 + \cdots .\label{gradact}\eeq

\begin{figure}
\centerline{\includegraphics[height=15cm,scale=1.0]{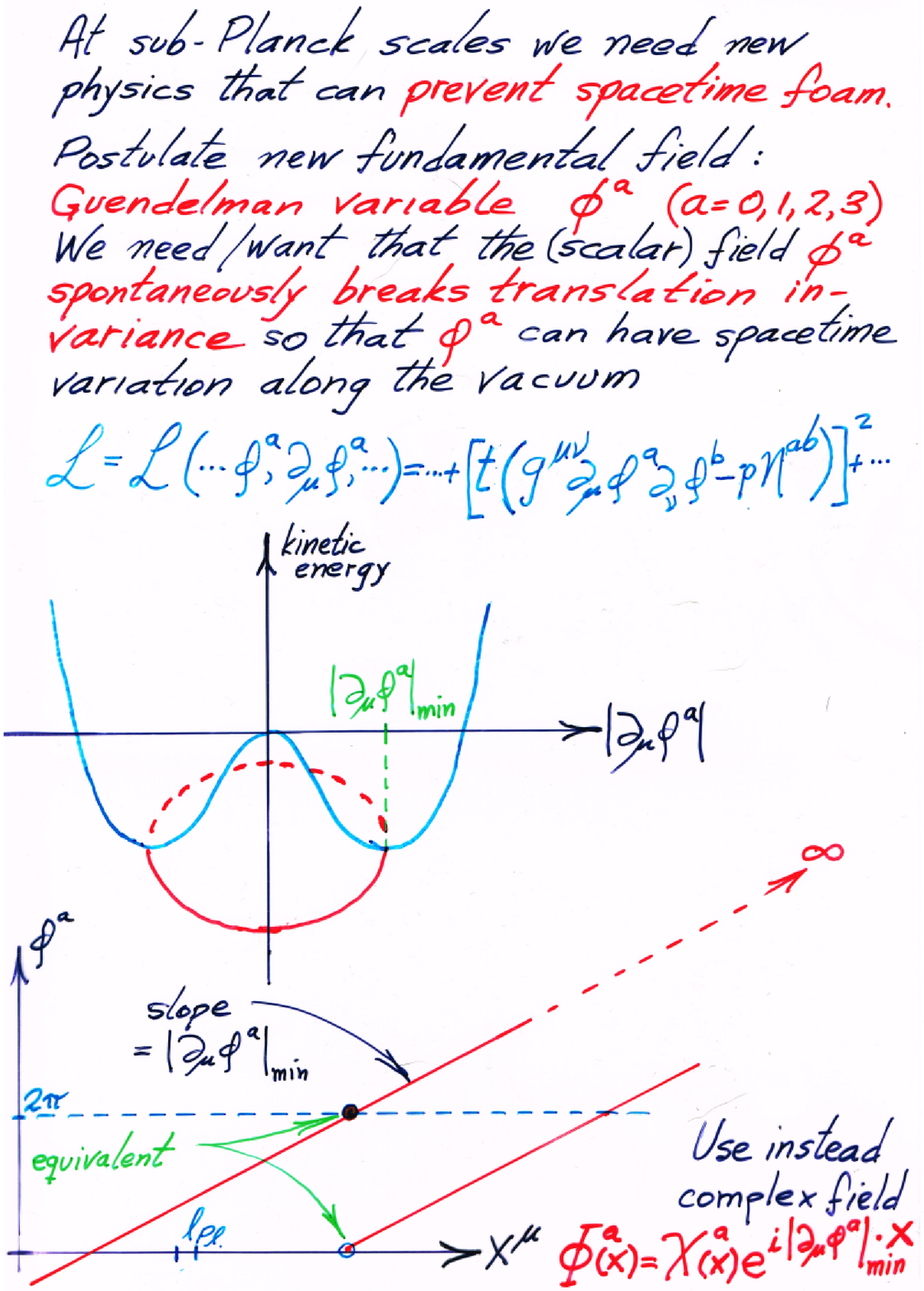}}
\caption{\label{fig1}At scales $a<<l_{Pl}$ we postulate a  new fundamental field $\Phi$
that explicitly breaks translational symmetry in the vacuum.}
\end{figure}

Such a contribution has terms quartic and quadratic in the
gradients that are respectively positive and negative and
therefore  result in a  ``Mexican hat potential'' as a function
of $\partial_{\mu}\phi^{a}$. The vacuum solution for such a
potential is a constant non-vanishing value
$|\partial_{\mu}\phi^{a}|_{min}$ of the gradient of the
fundamental fields (see Fig. \ref{fig1}). Such a vacuum
spontaneously breaks translational invariance of course.

Maintaining the  constant vacuum value
$|\partial_{\mu}\phi^{a}|_{min}$ for the gradient of $\phi^a$
in all of spacetime would lead to divergent values of the
fields $\phi^{a}$ at large distances. Therefore we take the new
fundamental fields to be the complex field $\Phi^{a}$
($a=0,1,2,3)$

\beq \Phi^a(x^0,x^1,x^2,x^3)\equiv
\Phi^{a}(x)=\chi^{a}(x)e^{i(\partial_{\mu}\phi^a)x}.\eeq

\nin For the moment we assume that the modulus $\chi^{a}(x)$
has the constant value $\chi_{0}$. In the vacuum it is the
gradient of the fundamental field $\phi^{a}(x)$ which has the
value $|\partial_{\mu}\phi^{a}|_{min}$ in the vacuum, i.e.,

\beq
\Phi^{a}_{vac}(x)=\chi^{a}_{0}e^{i(|\partial_{\mu}\phi^a|_{min})x}\eeq

\begin{figure}
\centerline{\includegraphics[scale=.5]{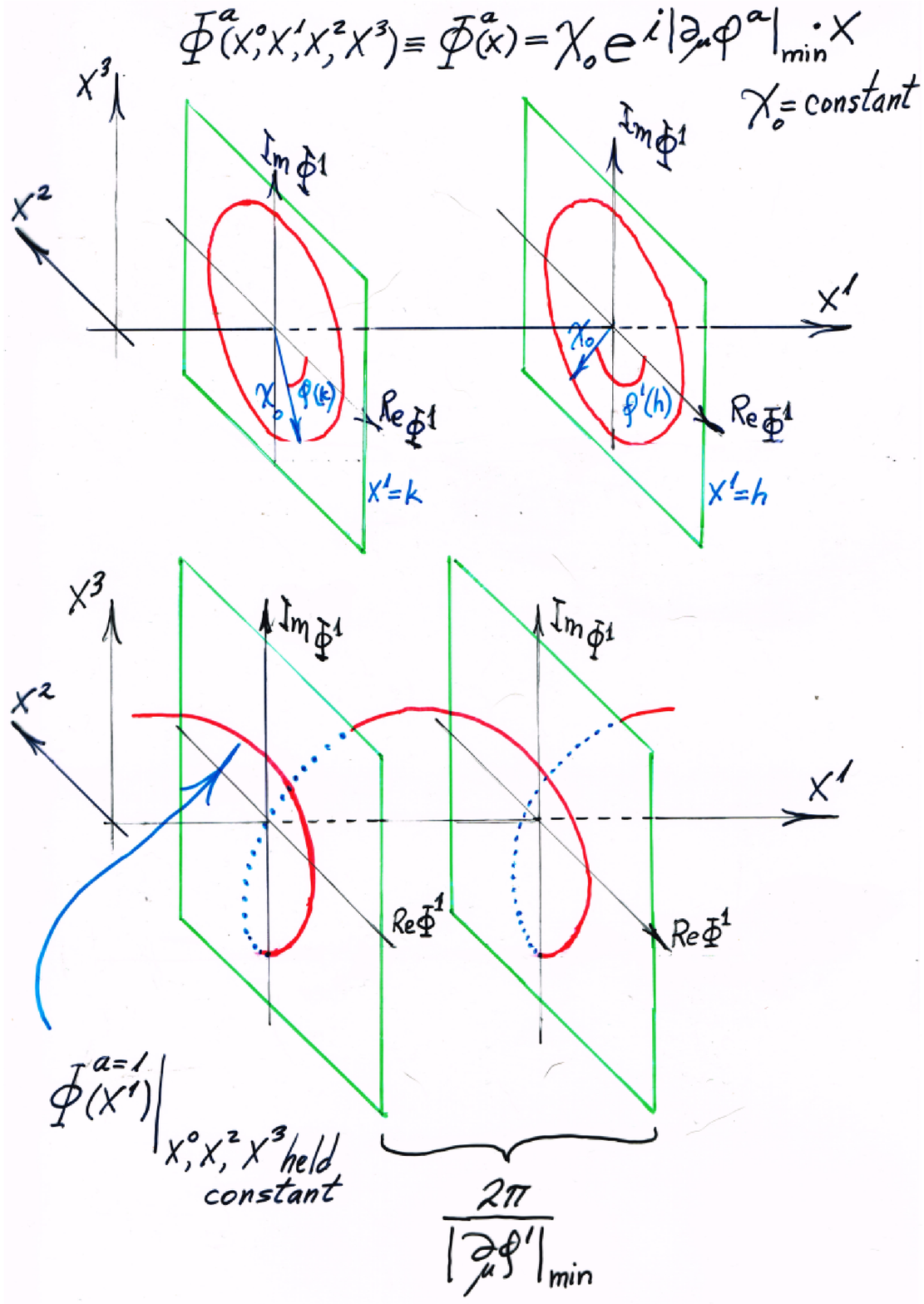}}
\caption{\label{fig2} The top part of the figure shows the value of the complex field $\Phi^1(x^1)$
at the arbitrary values $x^{1}=k$ and $x^{1}=h$. Really $\Phi^1(x^1)$ stands for
$\Phi^1(x^0,x^1,x^2,x^2)|_{x^0,x^2,x^3\mbox{held constant}}$  In the bottom part of the figure,
adjacent identical values
of $\Phi^{1}(x^1)$ define planes of constant $x^1$ separated by a distance $2\pi
/|\partial_{\mu}\phi^{a}|_{min}$.}
\end{figure}

We can also say that the condition for having the vacuum value
$|\partial_{\mu}\phi^{a}|_{min}$ for the gradient is that
planes corresponding to adjacent equal values of the complex
field $\Phi^{a}(x)$ are separated in spacetime by a (constant)
distance $2\pi /|\partial_{\mu}\phi^{a}|_{min}$. Fig.
\ref{fig2} shows the variation of the field component $\Phi^1$
as a function $x^1$

\begin{figure}
\centerline{\includegraphics[scale=.5]{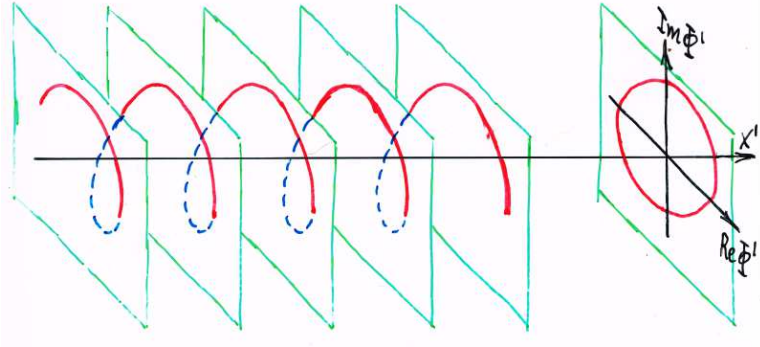}}
\caption{\label{fig3} Being in the ground state of the kinetic energy
potential of Fig. \ref{fig1} corresponds to the gradients of the fields
having the value $|\partial_{\mu}\phi^{a}|_{min}$ $a=0,1,2,3$ which
in turn corresponds to equidistant  planes with spacing proportional to
$2\pi /|\partial_{\mu}\phi^{a}|_{min}$. Here are shown several such planes for $a=1$.}
\end{figure}

\begin{figure}
\centerline{\includegraphics[scale=.5]{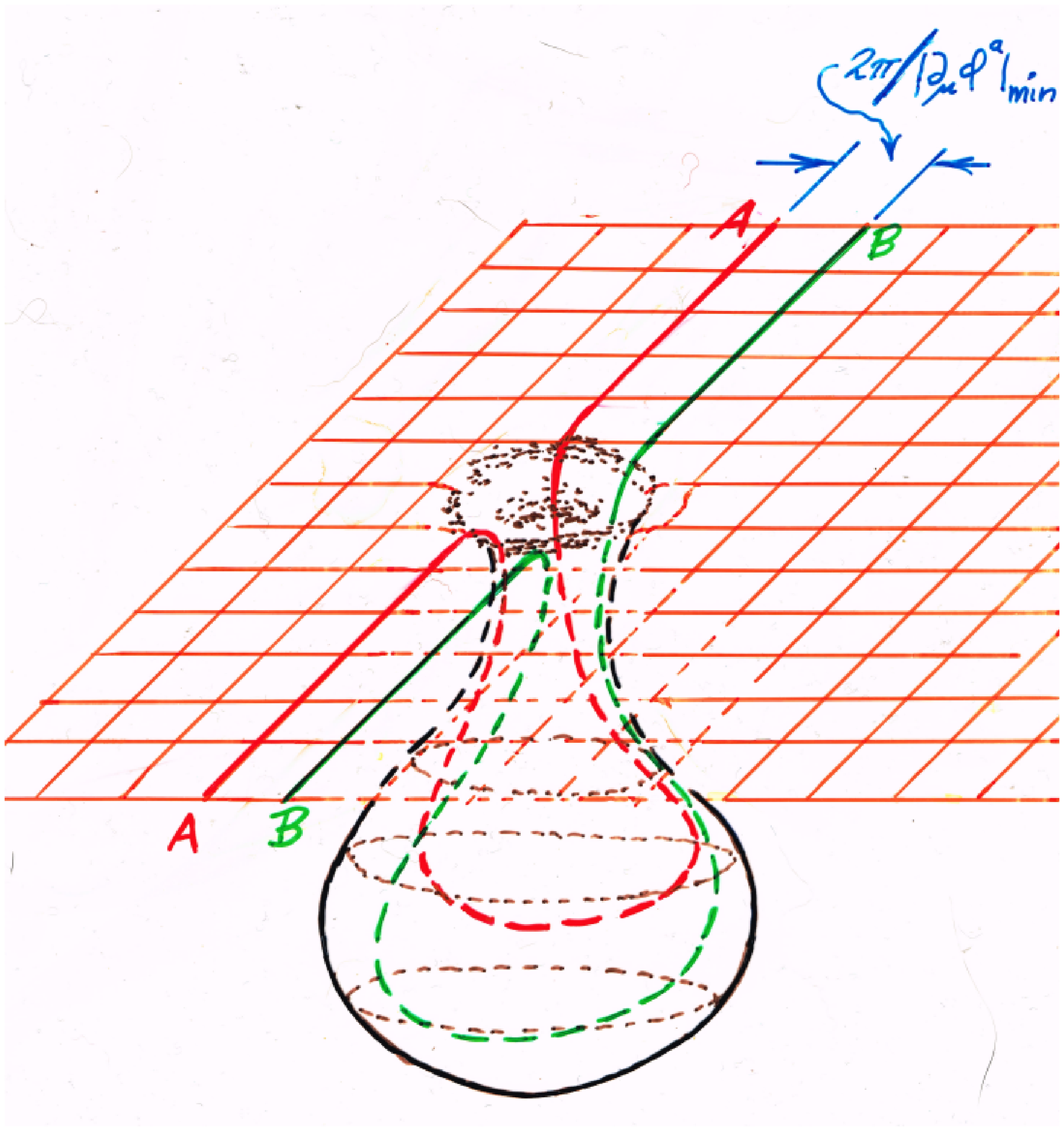}}
\caption{\label{fig4} Here we have an almost everywhere regular (i.e., flat spacetime) lattice
(here represented by a 2-dimensional grid drawn in perspective). The density of lattice
``planes'' corresponds to the vacuum value $|\partial_{\mu}\phi^a|_{min}$ of the gradient
of $\phi^a$ which equivalently means
that the distance between planes of some chosen constant value of $\Phi^a(x)$ is $2\pi
/|\partial_{\mu}\phi^{a}|_{min}$. However in the figure there is also an appendix (bubble)
that represents a departure from flat spacetime. If we follow two lattice ``planes'' $\mathbf{A}$ and
$\mathbf{B}$ in and out of the appendix we see that underway in the appendix the separation between these
``planes'' increases because by continuity the same number of lattice ``planes'' fill a larger volume of spacetime than
would be the case without the appendix (i.e., which would be just the ``volume'' corresponding to the appendix mouth).
Hence the density of lattice planes decreases in the appendix to a value $|\partial_{\mu}\phi^{a}|<
|\partial_{\mu}\phi^{a}|_{min}$ which corresponds to an excitation relative to  the vacuum state
(see Fig. \ref{fig1}). Departures from
flat spacetime costs energy.}
\end{figure}

So the requirement of being at the minimum of the potential in
Fig. \ref{fig1} (i.e., $
\partial_{\mu}\phi^a=|\partial_{\mu}\phi^{a}|_{min}$) defines
a (constant) density of planes each of which corresponds to the
same value of $\Phi^1$. Fig. \ref{fig3} shows a section of such
planes perpendicular to the $x^1$ axis.

There are similar planes for the other three spacetime axes.
Together this system of planes define a lattice with a lattice
constant equal to $2\pi /|\partial_{\mu}\phi^{a}|_{min}$
corresponding to the vacuum value for the gradients of the new
fundamental fields.

So we have seen that an action containing positive quartic and
negative quadratic terms in the gradient of the new proposed
fundamental fields $\phi^a$ (see \ref{gradact}) favours the
maintenance of a constant density of lattice points with
lattice constant $2\pi /|\partial_{\mu}\phi^{a}|_{min}$. Any
departure from this vacumm density of lattice points (or
planes) costs energy because it corresponds to moving away from
the minimum at
$|\partial_{\mu}\phi^a|=|\partial_{\mu}\phi^a|_{min}$.

This is the property that we need: an action that fixes the
density of lattice points in the sense explained above. In Fig.
\ref{fig4} we show a (locally 2-dimensional) appendix that
opens off of an otherwise  2-dimensional (flat space) lattice
with an almost everywhere fixed density of lattice points (or
planes) separated by the distance $2\pi
/|\partial_{\mu}\phi^{a}|_{min}$. By continuity the lattice
planes that go into the appendix must emerge again and rejoin
the flat spacetime lattice planes from which they originated.

The crucial point is that the appendix increases the ``volume''
of spacetime from that corresponding to area of mouth of the
appendix to the larger area of the interior of the appendix.
But by continuity the number of lattice planes entering and
leaving the appendix is the same as the number of planes that
enter and leave the area of the appendix mouth without the
appendix. Hence the  density of lattice planes (or points)
decreases within the appendix relative to the density within
the  area of the mouth without the appendix.

So the presence of the appendix relative to not having it
lowers the density of lattice points in the neighborhood of the
appendix. Within the appendix the lattice constant becomes
larger than $2\pi /|\partial_{\mu}\phi^{a}|_{min}$. This forces
the system away from the minimum at
$|\partial_{\mu}\phi^{a}|_{min}$ in the potential shown in Fig.
\ref{fig1}. Having the appendix costs energy. Energetically
flat spacetime is favoured.

\nin Notice that with an action of the form \ref{gradact} used
in our the pivotal relation Eqn. (\ref{pivotal}) is recovered
as an equation of motion upon taking a variation w.r.t.
$g^{\mu\nu}$

\section{The Emergence of Genearl Relativity as an Effective
Theory at Planck Scale}

When the new fundamental fields are introduced as the metric in
Eqn. (\ref{pivotal}) we have flat spacetime down to arbitrarily
small scales $a<< l_{Pl}$. And a consequence we have seen that
the Riemann curvature vanishes identically irrespective of what
action is used.

\begin{figure}
\centerline{\includegraphics[scale=.5]{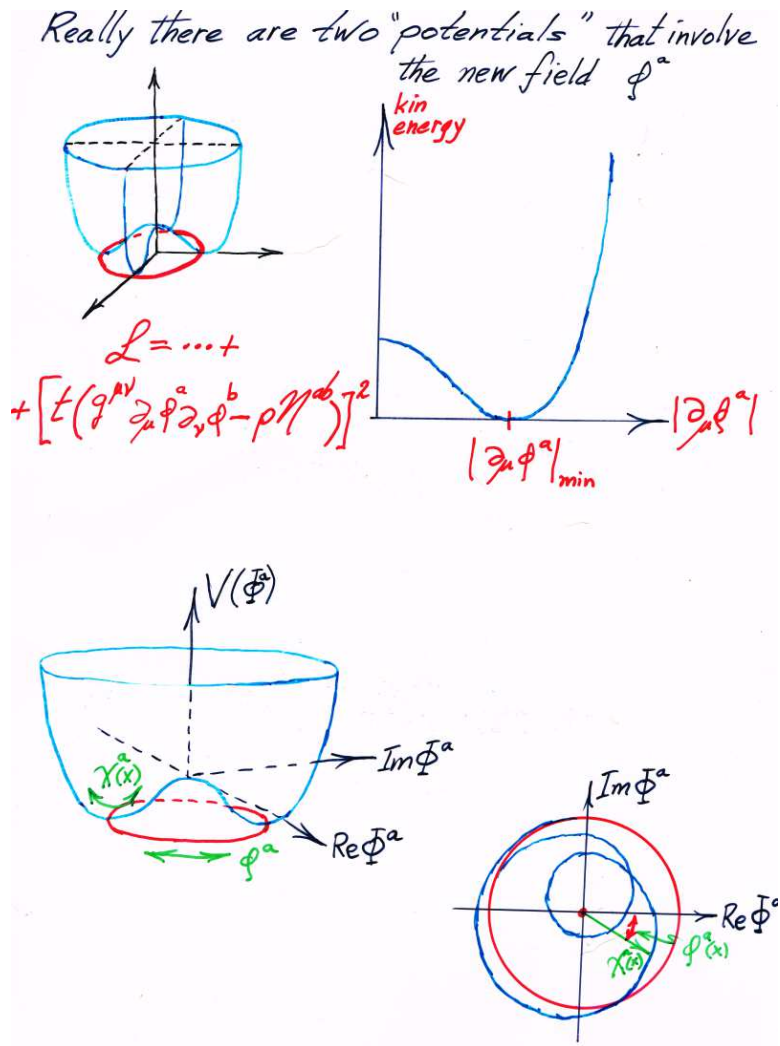}}
\caption{\label{fig5} The complex field has two degrees of freedom. In addition to the $\phi$-fields already discussed there
is also the $\chi (x)$. In the vacuum this degree of freedom is not excited and can be thought of as a soliton
with constant topology.}
\end{figure}

\begin{figure}
\centerline{\includegraphics[scale=.5]{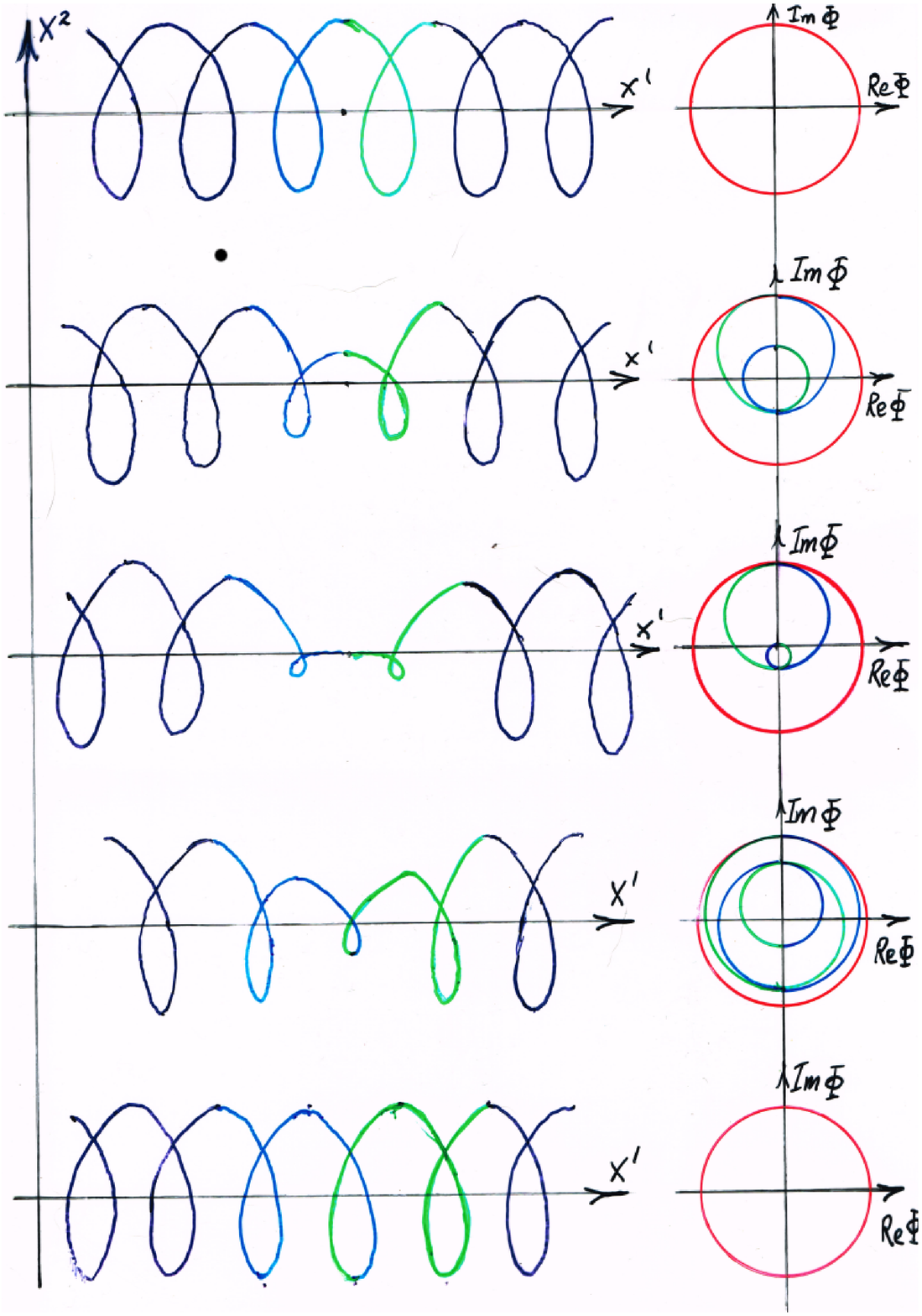}}
\caption{\label{fig6}If the $\chi (x) $ degree of freedom is sufficiently excited, the soliton can lose (or gain) a winding.
Thinking of the lattice discussed above, changes in the winding number for a soliton can be thought of as the
introduction of a crystal defect (dislocation line). It is known (see references to Hagen Klienert) that Einsteinian general relativity
can be formulated as a
``world crystal'' that has dislocation and disclination line defects that give rise to respectively torsion
and curvature. This presents a way that the usual general relativity can emerge as an effective theory at say the
Planck scale. Recall that at scales $a<<l_{Pl}$ where our new fundamental fields are important spacetime is
identically flat.
So phenomenologically we need a mechanism by which general relativity appears at roughly Planck scale.}
\end{figure}

 Now the question is how do we regain general
relativity when we go up to the Planck scale? Here we rely
heavily on the work\cite{Kleinert} of Hagen Kleinert. In the
special case of the model considered above we have seen how the
action defines a spacetime a lattice of constant density
$|\p_{\mu}\phi^a|_{min}$ consisting of planes corresponding to
equal values of the the new fundamental complex field
$\Phi^a(x)$. Now we think of this lattice as the ``world
crystal'' of Kleinert. Curvature (and torsion if desired) can
be introduced respectively as line dislocation and line
disclination defects. Fig. \ref{fig6} suggests in a soliton
model how a dislocation defect can come about by the loss of a
soliton winding. Kleinert demonstrates that the introduction of
disclination defects in a regular world crystal by the use
multivalued coordinate transformations reproduces general
relativity in full. We envision this happening at roughly the
Planck scale.

\FloatBarrier

\end{document}